\documentclass[apj,numberedappendix]{emulateapj}
\usepackage{amsmath}
\usepackage{hyperref}

\usepackage[english]{babel}
\usepackage{graphicx}
\usepackage{latexsym}
\usepackage{amsmath,amssymb}
\usepackage{natbib}
\usepackage{dcolumn}
\usepackage{bm}
\usepackage{calrsfs}
\usepackage{ulem}
\usepackage[usenames]{color}
\usepackage{xspace}

\newcommand{\ba}{\begin{eqnarray}}
\newcommand{\ea}{\end{eqnarray}}

\def\be{\begin{equation}}
\def\ee{\end{equation}}

\begin{document}

\title{Inhomogeneities in the light curves of Gamma-ray bursts afterglow}

\author{E.D.~Mazaeva$^{1}$,
A.S.~Pozanenko$^{1,2}$
and
P.Yu.~Minaev$^{1}$
}

\affil{
$^{1}$Space Research Institute, 84/32 Profsoyuznaya Str., Moscow 117997, Russia \\
$^{2}$National Research University Higher School of Economics, 20 Myasnitskaya Str., Moscow 101000, Russia \\
}

\begin{abstract}

We discuss the inhomogeneous behavior of gamma-ray burst afterglow light curves in optic.
We use  well-sampled light curves  based on mostly our own observations to find and identify deviations (inhomogeneities) from broken power law. By the inhomogeneous behavior we mean flashes, bumps, slow deviations from power law (wiggles) in a light curve.
In particular we report parameters of broken power law, describe phenomenology, compare optical light curves with X-ray ones and classify the inhomogeneities.
We show that the duration of the inhomogeneities correlates with their peak time relative to gamma-ray burst (GRB) trigger and the correlation is the same for all types of inhomogeneities.
\end{abstract}


\maketitle

\section{Introduction}
\label{sc:introduction}

In general, optical light curves of the gamma-ray bursts (GRBs) in the afterglow phase are described fairly well by smoothly broken-power law \citep{beu1999}
with temporal indices in the range from -0.5 to -2.5
\citep{kru2012}.
However, for the many well-sampled optical light curves after subtracting the host-galaxy component significant deviations (inhomogeneities) from the smoothly broken-power law are observed.
So far there is no clearly understanding of the physical processes whereby they are created.

There are several physical models explaining inhomogeneities.
One of the models suggests that the inhomogeneities are connected with the interaction of the fireball with moderate density enhancements in the ambient medium (density-jump model),
\citep{laz2002,dai2003}
but that density fluctuations are usually unable to produce either a sharp variation or a steep increase in the light curve, \citep{nak2003,nak2007}
for example, re-brightening of the GRB 030329 light curve at the first day
\citep{hua2007}.
Two-component jet model
\citep{lam2005}
with a narrow ultra-relativistic outflow and a wide but mildly relativistic ejecta for GRB 030329 well describes the radio data
\citep{ber2003,she2003}, however, the derived jet parameters indicated a wide jet, exhibiting the characteristic jet break about 10 days after the burst, which contradicts the observed jet break time of 0.5 days
\citep{lip2004}.
The most reliable is the energy-injection model
\citep{dai1998,zha2002}, when GRB fireball receives an additional energy injection  from  the  central  engine  during  the  afterglow phase,
but within this model it is difficult to explain the inhomogeneities at late stages of the afterglow (dozens of days).

In this paper we investigate the inhomogeneities of well-sampled optical and X-ray light curves of several GRBs.
\\
\\
\section{Observations and Data Sample}
\label{sc:obserations}

We analyze light curves of the GRB~030329, GRB~151027A, GRB~160131A, GRB~160227A and GRB~160625B.
A summary of the bursts' general properties is listed in Table~\ref{tab:GRBproperties}.
\begin{table}[]
\centering
\caption{Properties of the analyzed GRBs}
\begin{tabular}{ccccc} \hline
GRB	&	Trigger Time$^a$	&T$_{90}^{b}$ 	& Redshift 	&	Refs.\\
	&	(UT)	            &	(sec)		&		    &		 \\ \hline
030329\hphantom{0} &	11:37:14.67	&	22.9		&	0.1685 	& $^c$ \\
151027A		   &	03:58:24.15	&	130 $\pm$ 6	    &	0.81	& $^d$\\
160131A		   &	08:20:16.22	&	325 $\pm$ 72	&	0.972 	& $^e$\\
160227A		   &	19:32:08.09	&	317 $\pm$ 75	&	2.38 	& $^f$\\
160625B		   &	22:40:16.28	&	35.1 $\pm$ 0.2	&	1.406	& $^g$\\
\hline
\label{tab:GRBproperties}
\end{tabular}
\begin{flushleft}
$^{a}$ -- The time onset of the GRB ($=T_0$).\\
$^{b}$ -- For GRB 030329 T$_{90}$ is in the 30--400~keV band,
for others in the 15--350~keV band.\\
$^{c}$ -- \cite{van2004,gcn2020,gcn2053}\\
$^{d}$ -- \cite{gcn18478,gcn18496,gcn18487,gcn18493}\\
$^{e}$ -- \cite{gcn18951,gcn18959,gcn18965,gcn18966,gcn18969}\\
$^{f}$ -- \cite{gcn19098,gcn19106,gcn19109}\\
$^{g}$ -- \cite{gcn19581,zha2018}
\end{flushleft}
\end{table}

The optical data were obtained by Crimean Astrophysical Observatory (CrAO), Sayan Solar Observatory (Mondy), Tian Shan Astrophysical Observatory (TShAO), Maidanak High-Altitude Observatory,  Abastumani Astrophysical Observatory (AbAO), Special Astrophysical Observatory (SAO), ISON-Kislivodsk, ISON-Khureltogoot, ISON-NM observatories and taken from GCN observation report circulars\footnote{https://gcn.gsfc.nasa.gov/gcn3\_archive.html}.
The optical photometrical data reduction was based on IRAF\footnote{IRAF is distributed by the National Optical Astronomy Observatories 
}
standard tasks (/noao/digiphot/apphot).

\begin{figure}
\includegraphics[width=\columnwidth]{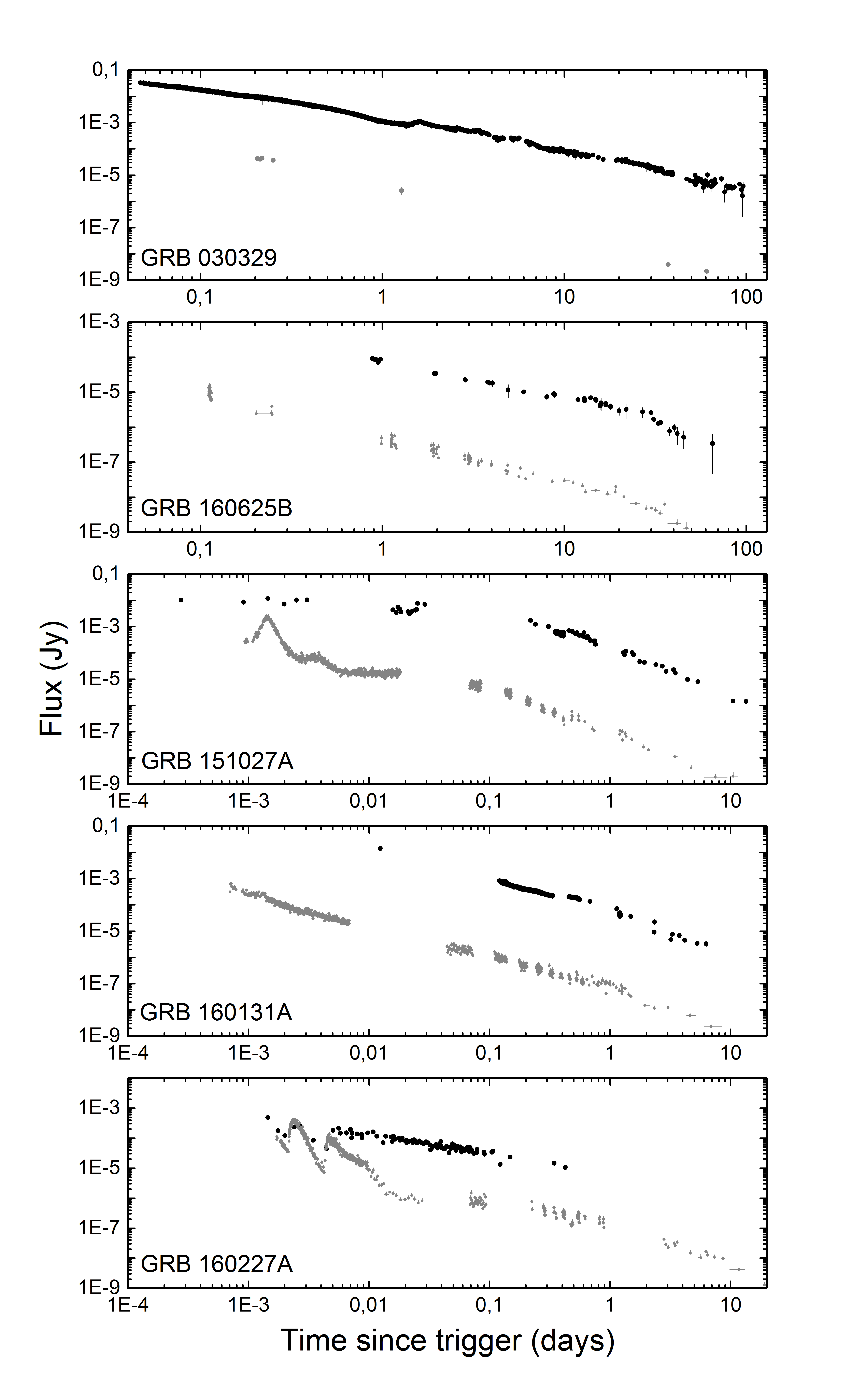}
\caption{The optical (black) and X-ray (gray) light curves of the analyzed GRBs, expressed in Jy.
X-ray band is 0.5--2~keV for GRB~030329 and 0.3--10~keV for other bursts.
The host galaxy contribution was subtracted in the light curves of GRB~030329 and GRB~160625B. }
\label{fig:opticalxray}
\end{figure}

A summary of optical observations of GRB~151027A, GRB~160131A and GRB~160227A will be presented in
\cite{maz2018},
optical data of GRB~030329 and GRB~160625B were taken from
\cite{lip2004,zha2018}.
The X-ray afterglow data of GRB~030329 were obtained by Rossi-XTE and XMM-Newton
\citep{tie2004}, the X-ray light curves of other GRBs received by Swift/XRT.\footnote{http://www.swift.ac.uk/xrt\_curves/}
The optical and X-ray light curves of the analyzed GRBs are presented in Fig.~\ref{fig:opticalxray}.

\section{Extraction of Inhomogeneities}
\label{sc:extraction}
The light curves of GRB~151027A, GRB~160131A and GRB~160625B were approximated by a smoothly broken power law (Beuermann function, see e.g. \cite{beu1999}):
\be
F = F_0\left[ \left( \frac{t-t_0}{t_{jb}}\right) ^{\alpha w} + \left( \frac{t-t_0}{t_{jb}}\right) ^{\beta w} \right] ^{-1/w} \,,
\label{eq:beuermann1}
\ee
where $\alpha$, $\beta$ are the early and late power law indices, $t_{jb}$ is time of jet-break, $w$ is the smoothing parameter.
The parameters $\alpha$, $\beta$, $t_{jb}$ were left free to vary, $t_{0}$ (time offset) and $w$ were fixed ($t_{0}$ = 0, $w$ = 1, 2, 3, 5, 7).

In case of GRB 160227A we use single power law model (\ref{eq:powerlaw}), as we have only data for the first day after the burst trigger, probably before a jet-break time for this burst
\be
F = F_0 t^{\alpha}\,.
\label{eq:powerlaw}
\ee
The GRB 030329 was modelled by a sum of two Beuermann functions
\be
F = \sum^2_{i=1} F_{0_i} \left[ \left( \frac{t}{t_{jb_i}}\right) ^{\alpha_iw_i} + \left( \frac{t}{t_{jb_i}}\right) ^{\beta_iw_i} \right] ^{-1/w_i} \,.
\label{eq:beuermann2}
\ee
The inhomogeneities (groups of points with significant deviation from the power-law-like behavior) in the optical light curve were excluded from the fitting procedure.

The best-fitting parameters are summarized in Table~\ref{tab:tablefit}.
\begin{table*}[]
\centering
\caption{The Fitting Parameters of Optical Light Curves}
\begin{tabular}{cccccccc} \hline
GRB	&	$i$	&	$F_0$ &	$w$	&	$t_{jb}$	&	$\alpha$ &	$\beta$		&	$\chi^{2}$/d.o.f.	\\
	&		&	(Jy)  &		&	(days)		&			 &				&		                \\\hline
030329\hphantom{0}&	1&(3.30 $\pm$ 0.02)$\times 10^{-5}$	&3 &0.524 $\pm$ 0.002&	--0.863 $\pm$ 0.001	&	--2.034 $\pm$ 0.006	&7053/3009	\\
	    &	2	&(4.86 $\pm$ 0.02)$\times 10^{-6}$	&2 &1.544 $\pm$	0.002	&\hphantom{0}7.327 $\pm$	0.084   &	--1.371 $\pm$ 0.005	&		\\
151027A	&		&(1.31 $\pm$ 0.15)$\times 10^{-4}$	&1 &0.193 $\pm$	0.016	&	--0.210 $\pm$ 0.046	&	--1.800 $\pm$ 0.034	&	256/83	\\
160131A	&		&(4.27 $\pm$ 1.66)$\times 10^{-4}$	&7 &1.223 $\pm$	0.314	&	--1.179 $\pm$ 0.008	&	--1.818 $\pm$ 0.156	&	333/497	\\
160227A &		&(7.44 $\pm$ 0.86)$\times 10^{-6}$	&--- &	---               &	--0.613 $\pm$ 0.024	&	---	                &	84/82	\\
160625B	&		&(2.42 $\pm$ 0.34)$\times 10^{-7}$	&5&	26.342 $\pm$ 3.112	&	--1.000 $\pm$ 0.017	&	--2.986 $\pm$ 0.767	&	94/91	\\
\hline
\label{tab:tablefit}
\end{tabular}
\end{table*}

\section{Classification of Inhomogeneities}
\label{sc:classification}
The optical light curves reveal a number of inhomogeneities, superposed over the power-law decay (see Table~\ref{tab:inhomogeneities}). The inhomogeneities were approximated by polynomials.

We separate the inhomogeneities into several classes. The classification is given in the following.
\begin{table*}[]
\centering
\caption{Fitting Parameters of Selected Optical Inhomogeneities}
\begin{tabular}{cccccccccc} \hline
GRB & $T_{peak}-T_{0}$$^{a}$& FWHM  & Amplitude$^{b}$ & Amplitude$^{b}$ & S/N$^{b}$ & Correlation  & Type$^{c}$ & Fig. \# \\
    & (days)                & (days)& (Jy)            & (mag)           & ($\sigma$)&   with X-ray &            &          \\\hline
030329\hphantom{0}	&	0.08826	&	0.03242	&	3.95	$\times 10^{-4}$&	0.02	&	7.5	    &	no  data	&	W (a)	&		\\
030329\hphantom{0}	&	0.14657	&	0.04640	& --3.20	$\times 10^{-4}$& --0.03	&	10.8	&	no  data	&	W (b)	&		\\
030329\hphantom{0}	&	0.23522	&	0.08627	&	2.12	$\times 10^{-4}$&	0.03	&	6.9	    &	no  data	&	W (c)	&		\\
030329\hphantom{0}	&	1.60190	&	0.24504	&	2.70	$\times 10^{-4}$&	0.32	&	57.9	&	no  data    &   n/c. &\ref{fig:inhomogeneities}(d)\\
030329\hphantom{0}	&	2.62084	&	0.14206	&	1.11	$\times 10^{-4}$&	0.21	&	19.2&	no  data	&	n/c.	&		\\
030329\hphantom{0}	&	3.39444	&	0.41709	&	1.36	$\times 10^{-4}$&	0.35	&	107.9&	no  data	&	n/c.	&		\\
030329\hphantom{0}	&	3.61772	&	0.33584	&	9.72	$\times 10^{-5}$&	0.28	&	44.3&	no  data	&	n/c.	&		\\
030329\hphantom{0}	&	5.67352	&	1.15475	&	8.52	$\times 10^{-5}$&	0.45	&	26.4&	no  data	&	n/c.	&		\\
030329\hphantom{0}	&	8.79820	&	0.41162	&	1.29	$\times 10^{-5}$&	0.15	&	13.1&	no  data	&	n/c.	&		\\
030329\hphantom{0}	&	9.76593	&	0.37217	&	1.83	$\times 10^{-5}$&	0.24	&	23.6&	no  data	&	n/c.	&		\\
030329\hphantom{0}	&	10.76485&	0.91008	&	1.59	$\times 10^{-5}$&	0.24	&	17.9&	no  data	&	n/c.	&		\\
030329\hphantom{0}	&	11.72378&	0.65981	&	1.14	$\times 10^{-5}$&	0.20	&	8.8	&	no  data	&	n/c.	&		\\
030329\hphantom{0}	&	12.58917&	0.41100	&	1.33	$\times 10^{-5}$&	0.26	&	8.1	&	no  data	&	n/c.	&		\\
151027A	&	0.53690	&	0.22988	&	2.17	$\times 10^{-4}$&	0.53	&	20.3&	yes		    &	F	&		\\
151027A	&	1.42918	&	0.25419	&	5.29	$\times 10^{-5}$&	0.48	&	10.8&	yes		    &	F	& \ref{fig:inhomogeneities}(a)\\
151027A	&	2.66541	&	1.41120	&	9.51	$\times 10^{-6}$&	0.38	&	6.6	&	no  data	&	n/c.	& \\
160131A	&	0.16648	&	0.04802	&	--1.76	$\times 10^{-5}$&	--0.04	&	7.1	&	no		    &	W (a)& \ref{fig:inhomogeneities}(c)\\
160131A	&	0.24792	&	0.05253	&	8.99	$\times 10^{-6}$&	0.03	&	5.7	&	no		    &	W (b)& \ref{fig:inhomogeneities}(c)\\
160131A	&	0.16764	&	0.01142	&	1.10	$\times 10^{-5}$&	0.03	&	3.1	&	no		    &	B	&		\\
160131A	&	0.52016	&	$>$0.05	&	5.77	$\times 10^{-5}$&	0.13	&	76.3&	no		&	B	&		\\
160227A	&	0.00190	&	0.00078	&	--2.15	$\times 10^{-4}$&	--1.06	&	29.4&	yes		&	n/c.	&		\\
160227A	&	0.00397	&	0.00145	&	--2.08	$\times 10^{-4}$&	--3.12	&	64.2&	yes		&	n/c.	&		\\
160625B	&	14.44238&	2.19765	&	1.40	$\times 10^{-6}$&	0.26	&	4.3	&	no		&	B	&\ref{fig:inhomogeneities}(b)	\\
\hline
\multicolumn{9}{l}{$^{a}$ Peak time ($T_{peak}$) of the inhomogeneities relative to the initial burst trigger ($T_0$).}\\
\multicolumn{9}{l}{$^{b}$ Amplitude and significance are relative to the power law fit of the light curve.}\\
\multicolumn{9}{l}{$^{c}$ B --- bump, F --- flare, W --- wiggle, n/c. --- nonclassified inhomogeneities.}\\
\label{tab:inhomogeneities}
\end{tabular}
\end{table*}
\subsection{Flares}
\label{sc:flares}
Flares (positive residuals) were first found in the X-ray light curve of GRB~970508
\citep{pir1998,pir2005},
later they have been observed in all phases of the canonical X-ray light curve
\citep{swe2013}.
In several GRB light curves, flares in X-ray and optical are synchronous.
In our sample we found two such X-ray/optical flashes in GRB 151027A (see Fig.~\ref{fig:inhomogeneities}(a)).
\subsection{Bumps}
\label{sc:bumps}
Variations with a positive residuals  and  without synchronous X-ray counterparts, we classified as the bumps (see Fig.~\ref{fig:inhomogeneities}(b)).
\subsection{Wiggles}
\label{sc:wiggles}
Wave-like variations with transition from positive to negative (and vice versa) residuals and small amplitudes (several millimags) of the early afterglow (up to approximately 0.5 days since GRB trigger) were detected in a dense photometric data for GRB 030329 and GRB 160131A (wiggles, see Fig.~\ref{fig:inhomogeneities}(c)).
\subsection{Nonclassified}
\label{sc:non-classified}
\begin{figure*}
\centering
\includegraphics[width=170mm,angle=-0]{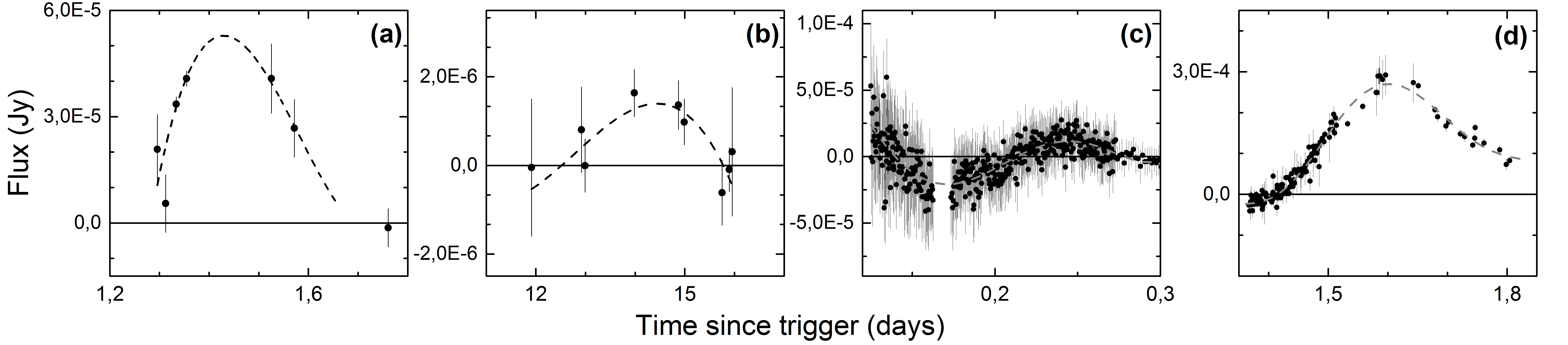}
\caption{Examples of inhomogeneities. The residuals over the power-law fits, representing inhomogeneities are shown.
(a) The optical flare of GRB 151027A;
(b) the optical bump of GRB 160625B;
(c) the optical wiggles of GRB 160131A;
(d) the nonclassified bump of GRB 030329.}
\label{fig:inhomogeneities}
\end{figure*}

Bumps with no synchronous detection in X-rays because of the absence of the corresponding X-ray data (like inhomogeneities of GRB 030329) or the inhomogeneities that do not fit the classification criteria are named as nonclassified (see Fig.~\ref{fig:inhomogeneities}(d)).
In particular, we could not classify the inhomogeneities of the GRB 160227A --- the optical light curve has a complex structure with an additional component, which is not visible in X-rays.

\section{The FWHM - T$_{\rm peak}$ Relation}
\label{sc:fwhm-tpeak}

\begin{figure*}
\centering
\includegraphics[width=130mm,angle=-0]{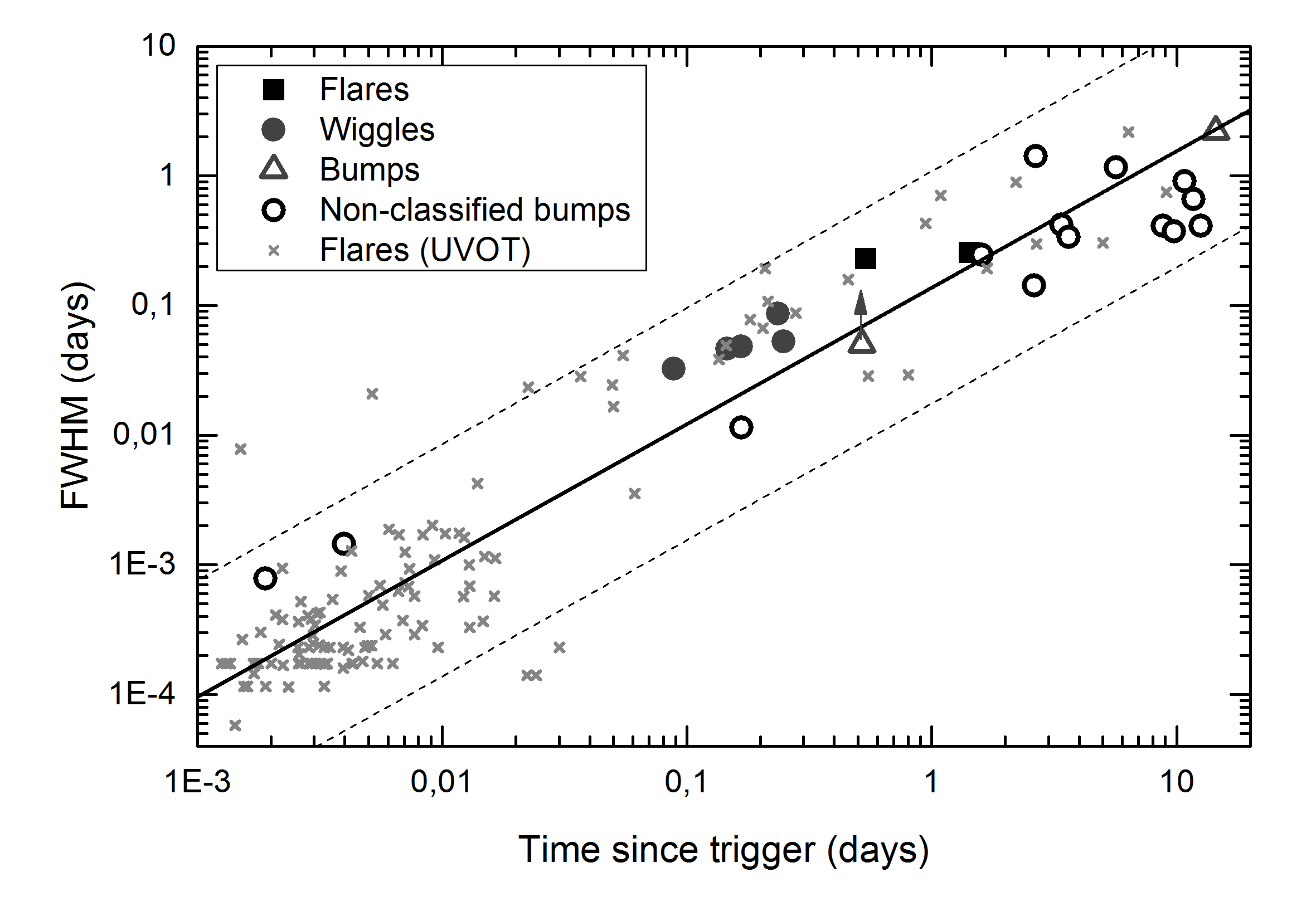}
\caption{The FWHM - T$_{\rm peak}$ relation for inhomogeneities, constructed for ones from our sample and for flares detected by UVOT/Swift from \cite{swe2013}. Thick solid line represents power-law fit to the joint sample, dotted lines bound 2 sigma correlation region.}
\label{fig:FWHMtime}
\end{figure*}
For found parameter of FWHM and T$_{\rm peak}$ we constructed scatterplot, presented in Fig.~\ref{fig:FWHMtime}.
The optical flares detected by UVOT/Swift in
\cite{swe2013}
are also plotted at Fig.~\ref{fig:FWHMtime}.
For the UVOT sample only start and stop times of flashes are available, i.e. total duration of the flares.
We use  half of duration for each flash to put it on the  Fig.~\ref{fig:FWHMtime}.
The correlation between FWHM and T$_{\rm peak}$ found previously in
\cite{yi2017}
is evident.
We fitted the FWHM - T$_{\rm peak}$ scatterplot for the combined sample of FWHM and half time using the power-law logarithmic model: {$\log\left(\frac{FWHM}{1~ \rm day}\right) = (1.05 \pm 0.03) \log \left(\frac{T_{\rm peak}}{1~ \rm day}\right) + (-0.86 \pm 0.07)$.
Power-law index of $\simeq$ 1 indicates the linear dependence of the investigated parameters: FWHM $\sim$ T$_{\rm peak}$.
Earlier, the positive correlation between the arrival time and duration of X-ray flares was noted in
\cite{per2005}.
It is interesting that all types of  inhomogeneities introduced previously (wiggle, flare, bump, etc.) follow the same correlation (see Fig.~\ref{fig:FWHMtime}), possibly indicating their similar physical nature.

\section{Discussion}
\label{sc:discussion}
\begin{figure}
\includegraphics[width=\columnwidth]{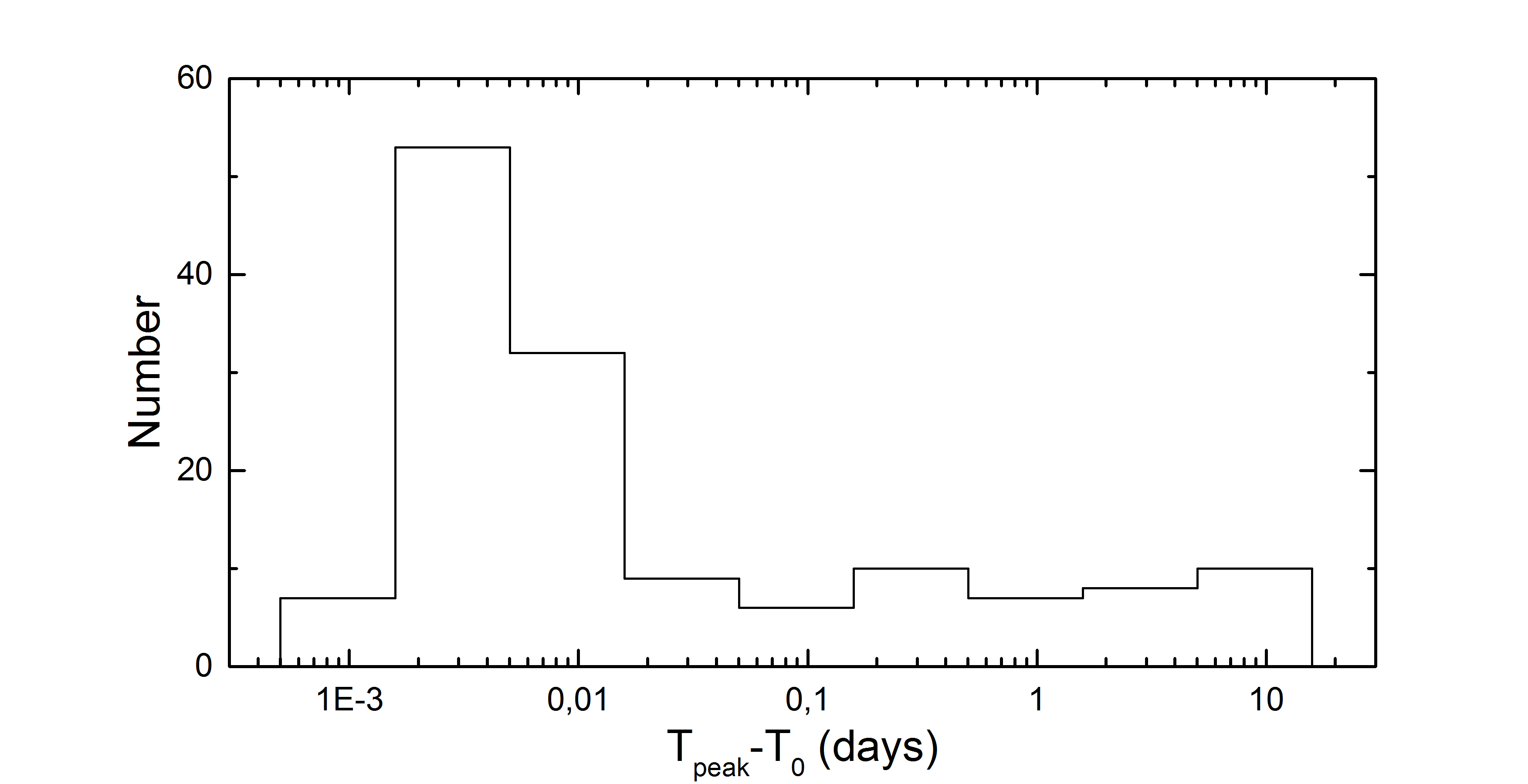}
\caption{Number density of a peak time of  inhomogeneities in optic light curves of GRB afterglow.}
\label{fig:NumberTpeak}
\end{figure}

In this paper, we are analyzing the inhomogeneities of GRB optical light curve afterglow. There are totally 23 inhomogeneities  identified in five well sampled light curves of GRBs. The inhomogeneities were classified to the following types: flares, bumps, wiggles, etc.
The sample of 119 UV/optical flares from
\cite{swe2013},
mostly observed at early times (T$_{peak}$ $<$ 0.02 days) was jointly analyzed.
We add to the sample late time inhomogeneities (21 at T$_{peak}$ $>$ 0.08 days).

All types of inhomogeneities from our sample and UVOT flares follow the same correlation between FWHM and T$_{peak}$, suggesting similar physical nature or strong selection effect. The power law index of the dependence is about 1 indicating a linear dependence of FWHM and T$_{peak}$.

This dependence differs from the case of prompt emission when there is no correlation of the pulse duration with the arrival time
\citep{mit1988}.

One can suggest a gap in a number density of inhomogeneities between 0.02~$<$~T$_{peak}$~$<$~0.04 days in Fig.~\ref{fig:FWHMtime}. However, the peak time distribution of the inhomogeneities reveals no signatures of the gap or any kind of bimodality (see Fig.~\ref{fig:NumberTpeak}).



\end{document}